\documentclass[12pt]{aastex}
\usepackage{amsmath}
\newif\ifAMStwofonts
\AMStwofontstrue

\def\hi{\mbox{\rm H{\sc~i}}}

\def\arcdeg{\hbox{$^\circ$}}
\def\slantfrac#1#2{\hbox{$\,^#1\!/_#2$}}

\def\nl{\cr}

\def\ga{\mathrel{\hbox{\rlap{\hbox{\lower4pt\hbox{$\sim$}}}\hbox{$>$}}}}
\def\la{\mathrel{\hbox{\rlap{\hbox{\lower4pt\hbox{$\sim$}}}\hbox{$<$}}}}
\defcitealias{fw90}{FW90}
\defcitealias{vlm10}{VLM10}
\vspace{3cm}
\shorttitle{Pulsar Distances}
\shortauthors{J. P. W. Verbiest  et al.}

\begin{document}
\title{On Pulsar Distance Measurements and their Uncertainties}

\author{J.~P.~W.~Verbiest} 
\affil{Max-Planck-Institut f\"ur Radioastronomie, Auf dem H\"ugel 69,
  53121 Bonn, Germany}

\author{J.~M.~Weisberg and A.~A.~Chael}
\affil{Department of Physics and Astronomy, Carleton College,
  Northfield, MN 55057, USA}

\author{K.~J.~Lee}
\affil{Max-Planck-Institut f\"ur Radioastronomie, Auf dem H\"ugel 69,
  53121 Bonn, Germany}

\author{D.~R.~Lorimer} 
\affil{Department of Physics, West Virginia University, Morgantown, WV
  26506, USA}

\begin{abstract}
Accurate distances to pulsars can be used for a variety of studies of
the Galaxy and its electron content. However, most distance measures
to pulsars have been derived from the absorption (or lack thereof) of
pulsar emission by Galactic \hi\ gas, which typically implies that
only upper or lower limits on the pulsar distance are available. We
present a critical analysis of all measured \hi\ distance limits to
pulsars and other neutron stars, and translate these limits into
actual distance estimates through a likelihood analysis that
simultaneously corrects for statistical biases. We also apply this
analysis to parallax measurements of pulsars in order to obtain
accurate distance estimates and find that the parallax and
\hi\ distance measurements are biased in different ways, because of
differences in the sampled populations. Parallax measurements
typically \emph{under}estimate a pulsar's distance because of the
limited distance to which this technique works and the consequential
strong effect of the Galactic pulsar distribution (i.e. the original
Lutz-Kelker bias), in \hi\ distance limits, however, the luminosity
bias dominates the Lutz-Kelker effect, leading to \emph{over}estimated
distances because the bright pulsars on which this technique is
applicable are more likely to be nearby given their brightness.
\end{abstract}
\keywords{astrometry, pulsars}

\section{Introduction}\label{sec:intro}
The rotation of pulsars, which causes their continuous emission to be
observed as highly regular pulses, makes these objects highly useful
probes of any dispersive phenomena in interstellar space. Combined
with an accurate and precise distance, pulsar emission (specifically
its dispersion and Faraday rotation) provides crucial information for
modelling of the Galactic electron distribution and magnetic field.

Parallax measurements are non-trivial undertakings and only very few
significant parallax measurements \citep{gtwr86,bmk+90a} were made
within the first two decades after pulsars were discovered. Another
method to determine a pulsar's distance is based on Galactic
\hi\ spectra in the direction to the pulsar. This method (known as the
\emph{kinematic} or \hi\ method) compares the \hi\ spectrum on-pulse
(when the pulsar emission is seen) and off-pulse (when the pulsar
emission beam is turned away). Any observed pulsar absorption must
originate in gas lying closer than the pulsar; while gas located
farther than the pulsar will \emph{not} exhibit absorption.  The
velocities of these respective \hi\ regions are subsequently derived
from the spectrum and translated to distances with help of a Galactic
rotation model. The distance of the furthest \hi\ gas that appears in
absorption then provides a lower limit $D_{\rm low}$ on the pulsar
distance, while the distance of the nearest gas that only appears in
emission, is interpreted as an upper limit $D_{\rm up}$ on the pulsar
distance.

Roughly two decades after the discovery of pulsars, \citet[henceforth
  FW90]{fw90} collated all published pulsar distances, which at the
time consisted of 50 \hi\ distances, three parallax measurements and
20 distances by association. Given the importance of \hi\ distances,
they critically investigated the various measurements and defined a
set of criteria that has been used in almost all subsequent
publications.

Progress in both interferometric hardware (at the Long Baseline Array
in the South and the Very Long Baseline Array in the North) and in the
sensitivity of pulsar timing, subsequently allowed an exponential
increase in the number of measured pulsar parallaxes so that currently
57 parallaxes are measured. This led \citet[henceforth VLM10]{vlm10}
to collate those distances and investigate the statistical bias
predicted by \citet{lk73}. The work presented by \citetalias{vlm10} was
based on a Bayesian analysis that took into account both the Galactic
distribution of pulsars (which is the actual bias first discussed by
Lutz and Kelker in 1973\nocite{lk73}) and the intrinsic pulsar
luminosity distribution; but they only considered parallax
measurements.

In this paper, we present an update of the work done by
\citetalias{fw90}: we list all 80 published distances to pulsars and
other neutron stars, based on \hi\ measurements or associations with
objects having \hi\ distances, and evaluate them based largely on the
criteria laid out by \citetalias{fw90}. We then improve the analysis
of \citetalias{vlm10} by deriving fully analytic solutions that
replace the need for (approximate) Monte-Carlo simulations. Also, the
\citetalias{vlm10} analysis is expanded to incorporate information
provided by \hi\ distance limits; and to provide bias-corrected
\emph{distances} in addition to parallaxes. As in the case of
\citetalias{vlm10}, the present paper bases its bias-correction method
on empirical models for the Galactic pulsar distribution and
luminosity function. These models do add an unquantified level of
uncertainty to the analysis, but can easily be updated as our
knowledge about the pulsar population grows through pulsar
surveys. The evaluation of \hi\ distance limits is presented in
Section~\ref{sec:HID}; the likelihood analysis to correct for the
biases is derived in Section~\ref{ssec:Bayes}. Bias-corrected
parallaxes and distances are given in tables~\ref{tbl:dist} and
\ref{tbl:vlm10} and a summarising discussion is found in
Section~\ref{sec:results}.

%

\section{\hi\ Kinematic Distances}\label{sec:HID}

\subsection{Source Selection and \hi\ Kinematic Distance Limit Determination}
\label{ssec:velocity}
\citetalias{fw90} established standard techniques for the extraction of
reliable pulsar kinematic distance limits. Specifically, they defined
the bound of $T_{\rm b} \geq 35\,$K on the brightness temperature of
\hi\ emission used for deriving upper distance limits\footnote{They
  also note this bound can be relaxed depending on the sensitivity of
  the observation, provided the optical depth for the emission is 0.3
  or higher.}, pointing out that weaker emission would not be expected
to result in significant absorption. Secondly, they re-evaluated
distance limits based on old Galactic models and rotation curves,
defaulting to the IAU values for the distance of the Solar System to
the centre of the Galaxy [${\rm R_0} = 8.5\,$kpc] and the Galactic
rotation velocity in the Solar System neighborhood [$\Theta_0 =
  220\,$km~s$^{-1}$]\citep{klb86}, and using the flat rotation curve
of \citet{fbs89}. In converting velocities to distances, furthermore,
they assumed an uncertainty of 7\,km~s$^{-1}$ because of known random
motions of that order \citep{dl90}. Finally, in the Perseus arm, with
its well-known spiral shock, they either used independent distance
tracers or applied the approximation proposed by \citet{jrd89}, which
states the global rotation curve can be applied (near $G_{\rm l} =
130^{\circ}$) provided the measured \hi\ velocities are decreased by a
factor of 1.6. Most investigations since then have used these same
criteria and so does the present paper, with a few exceptions as
listed below.
%

In the current work, we present a uniformly-determined sample of
neutron star \hi\ kinematic upper and lower distance estimates by
finding all such efforts in the literature, and then applying the
\citetalias{fw90} criteria to any published data that have not
previously been analyzed with that procedure. If the cited authors
made a good case for a non-flat rotation curve (e.g, in the direction
of the Galactic bar, the 3 kpc arm, or the Perseus arm shock), we
maintain their curve in our analysis. If, however, the original
authors used a flat rotation curve but non-IAU Galactic constants, we
reanalyze the kinematic distance limits, using the flat rotation curve
{\it{and}} IAU constants. We note that \citet{rmz+09} obtain a
Galactic rotation velocity that is larger than the IAU value at a
signficance of 95\%. More recent measurements by the same authors have
increased the significance of this offset to close to 99\% (Reid,
personal communication). Based on Eq. 2.21 of \citet{dt91}, we find
that this could imply an overestimate of our \hi\ distance limits by
up to $\sim 20 \%$ (though generally much less), depending on Galactic
longitude and the measured \hi\ velocity. 
%
%

In Table \ref{tbl:dist}, we list the \hi\ kinematic lower and upper
distance limits, $D_{\rm low}$ and $D_{\rm up}$ respectively; and
their uncertainties, $\sigma_{\rm low}$ and $\sigma_{\rm up}$, to
conventional radio pulsars as well as to otherwise radio-quiet neutron
stars with radio bursts; and to supernova remnants (SNRs) securely
associated with various kinds of neutron stars. The values shown are
from the stated authors' work, unless otherwise indicated in the
table. If the original authors gave distance limits meeting our
criteria, but neglected to derive uncertainties on these limits, then
we do so ourselves according to the procedure laid out in
\citetalias{fw90}.  In such cases, the table entry's reference shows a
superscript $a$. If we judge that the listed authors' distance limits
themselves need adjustment, we do so and mark the entry's reference
with a superscript $b$, and describe details of any such changes in \S
\ref{sec:individsources}. Two sources, marked with the superscript
$f$, had two or more original sets of distance limits because the
cited authors evaluated multiple Galactic rotation models without
expressing a clear preference for one; in those cases we select the
one using the standard flat rotation curve, for overall
consistency. (Note that in none of these cases the various rotation
models provided significantly different results.)

\subsection{Notes on individual sources}\label{sec:individsources}
In the subsections below, we explain any adjustments to criteria that
led to the originally published upper and/or lower distance
limits. The values themselves are summarized in Table \ref{tbl:dist}.

\subsubsection{SNR Kesteven 73 and AXP 1E1841-045}
\citet{tl08a} performed an \hi\ kinematic distance study of the
supernova remnant Kes 73, which is associated with AXP 1E1841-045.
The authors showed that the SNR absorption extends to the tangent
point, 7.5 kpc distant, which marks the lower distance limit. They
also made the case that the lack of absorption at $v = 84$ km s$^{-1}$
sets an upper distance limit on the far side of the tangent point.  We
find these arguments compelling.  However, we find that the flat
rotation curve then indicates that $D_{\rm up} = 10.2$\,kpc, whereas
\citet{tl08a} quoted $D_{\rm up} = 9.8$\,kpc for a flat rotation curve.
 
\subsubsection{PWN G54.1+0.3 and PSR J1930+1852}
\citet{ltw08} analyzed \hi\ spectra of PWN G54.1+0.3, which is
associated with PSR J1930+1852. We confirm that the lower distance
limit is at the tangent point. While they place the upper distance
limit at the Solar Circle on the far side of the Galaxy due to a lack
of any negative velocity absorption, we instead adhere to the
procedure of \citetalias{fw90}, relaxing the limit to the distance
corresponding to the first strong emission at negative velocities not
showing absorption, i.e., at $v = -30$\,km s$^{-1}$.  After resetting
the rotation curve to the flat model with IAU galactic constants, we
then find that $(D_{\rm low},D_{\rm up}) = (5.0,12.6)$\,kpc.

\subsubsection{SNR CTB 80 and PSR B1951+32}
\citet{ss00} measured the \hi\ absorption spectrum of SNR CTB 80,
which is associated with PSR B1951+32. There is significant absorption
out to the tangent point, yielding $D_{\rm low} = 3.1$\,kpc.  Unfortunately,
the published absorption spectrum does not extend below $v = -15$\,km
s$^{-1}$, which is insufficient to establish a $D_{\rm up}$
measurement.

%
%

\section{Lutz-Kelker Bias and Corrections}\label{ssec:Bayes}
\citet{lk73} first presented the argument that because of the
non-linearity of sample volume with distance, objects are
statistically more likely to be further away rather than closer
by. Correction for this bias \citepalias[which is related to the
  Malmquist bias but is more correctly named Lutz-Kelker bias, as
  discussed by][]{vlm10} is relatively straightforward through a
likelihood analysis that incorporates probabilities derived from a
variety of possible measurements. Our derivation is similar to that of
\citetalias{vlm10} but differs in a few fundamental areas. First, the
primary focus of \citetalias{vlm10} was biases in parallax
measurements, while our analysis considers both parallax and distance,
which is a more natural quantity when dealing with \hi\ distance
limits. (Note that the conversion between parallax and distance is not
a simple inversion in the case of finite uncertainties, as the
transformation between these two quantities is non-linear.)  Second,
where \citetalias{vlm10} applied a Bayesian analysis with prior
information based on the pulsar luminosity and position in the Galaxy,
we consider these quantities as measurements and have hence no need
for prior information at all, which removes the Bayesian character of
this analysis and leaves a straightforward likelihood
analysis. Effectively this is no more than an aesthetic difference,
however, which does not affect the results. Indeed, our approach could
be considered Bayesian with a uniform prior. In particular, our
analysis considers the following possible measurements:
\begin{itemize}
\item a parallax measurement, $\varpi_{\rm meas}$;
\item a lower \hi\ distance limit, $D_{\rm low}$;
\item an upper \hi\ distance limit, $D_{\rm up}$;
\item the pulsar radio flux, $S$ (measured at or near an observing
  frequency of $1.4\,$GHz);
\item and the pulsar's Galactic position, $G_{\rm l}, G_{\rm b}$.
\end{itemize}
Given a subset or all of these measurements and assuming no
correlations between these values, we can determine the probability
density function of the pulsar distance, $D$, through
\begin{equation}\label{eq:Basic}
p\left(D\vert\varpi_{\rm meas}, D_{\rm low}, D_{\rm up}, S, G_{\rm l},
G_{\rm b}\right) = 
p\left(D\vert\varpi_{\rm meas}\right)
p\left(D\vert D_{\rm low}\right)
p\left(D\vert D_{\rm up}\right)
p\left(D\vert S\right)
p\left(D\vert G_{\rm l}, G_{\rm b}\right).
\end{equation}
In the above equation (as in all equations throughout this paper), we
only explicitly state dependence on parameters, while dependence on
the uncertainties of said parameters is implied. In other words, where
we write $p(D\vert\varpi_{\rm meas})$, we really mean
$p(D\vert\varpi_{\rm meas}, \sigma_{\varpi_{\rm meas}})$. In the
following, these five terms will be derived; they will respectively be
referred to as the parallax term, the lower \hi\ limit term, the upper
\hi\ limit term, the luminosity term and the volumetric or Galactic
term.

\subsection{The Parallax Term, $p\left( D \vert \varpi_{\rm meas} \right)$}
Given a measurement $\varpi_{\rm meas}$ with uncertainty
$\sigma_{\varpi}$ and assuming a Gaussian uncertainty distribution,
the probability of the true parallax given the data is
\begin{equation}
p\left( \varpi \vert \varpi_{\rm meas} \right) \propto
\frac{1}{\sqrt{2 \pi} \sigma_{\varpi}}
\exp\left[-\frac{1}{2}\left(\frac{\varpi_{\rm meas} - \varpi}
  {\sigma_{\varpi}}\right)^2\right].
\end{equation}
Since $p\left( D \right) = \vert \partial \varpi / \partial D
\vert p\left( \varpi \right) \propto p\left( \varpi \right) / D^2$,
this means
\begin{equation}\label{eq:pxTerm}
  p\left( D \vert \varpi_{\rm meas} \right) \propto
  \frac{1}{D^2}
\exp\left[-\frac{1}{2}\left(\frac{\varpi_{\rm meas} - 1/D}
  {\sigma_{\varpi}}\right)^2\right].
\end{equation}

In the case of asymmetric uncertainties on parallax measurements
\citep[as given, e.g., by][]{cbv+09}, we assume 
\begin{equation}\label{eq:AsymPx}
  p\left( \varpi \vert \varpi_{\rm meas}\right)
  \propto
  H\left(\varpi-\varpi_{\rm meas}\right) \exp\left[-\frac{1}{2} \left(
    \frac{\varpi_{\rm meas}-\varpi}{\sigma_{\varpi {\rm
          up}}}\right)^2\right]
  + H\left(\varpi_{\rm meas}-\varpi\right) \exp\left[-\frac{1}{2}
    \left( \frac{\varpi_{\rm meas}-\varpi}{\sigma_{\varpi {\rm
          low}}}\right)^2\right],
\end{equation}
with $\varpi+\sigma_{\varpi {\rm up}}$ and $\varpi-\sigma_{\varpi {\rm
    low}}$ respectively the upper and lower limit of the 1\,$\sigma$
interval of the measurement's probability density function; and with
$H(x)$ the Heaviside step function, for which
\begin{equation}
H(x)=
\begin{cases}
0 & \textrm{if } x<0,\\
0.5 & \textrm{if } x=0,\\
1 & \textrm{if } x>0.
\end{cases}
\end{equation}
For distance, as in the symmetric case, the extra factor of $D^{-2}$
is added, resulting in
\begin{align}\label{eq:pxAsymTerm}
p\left( D \vert \varpi_{\rm meas} \right) \propto &
\frac{1}{D^2}
H\left(1/D-\varpi_{\rm meas}\right) \exp\left[-\frac{1}{2}\left(
  \frac{\varpi_{\rm meas}-1/D}{\sigma_{\varpi {\rm
        up}}}\right)^2\right]\nonumber\\
 & + \frac{1}{D^2}
H\left(\varpi_{\rm meas}-1/D\right) \exp\left[-\frac{1}{2}\left(
  \frac{\varpi_{\rm meas}-1/D}{\sigma_{\varpi {\rm low}}} \right)^2
  \right].
\end{align}

\subsection{The \hi\ Distance Limit Terms, $p \left( D \vert D_{\rm
  up}, D_{\rm low}\right)$}\label{sss:HI}
Assuming the distance of the furthest absorbing \hi\ gas is determined
to be $D_{\rm low}$ with measurement uncertainty $\sigma_{\rm low}$,
then the probability distribution of the actual distance of the
limiting gas is given (assuming Gaussian uncertainties) by
\begin{equation}
p\left( d \vert D_{\rm low}\right) \propto
\frac{1}{\sqrt{2\pi}\sigma_{\rm low}} \exp \left[ -\frac{1}{2} \left(
  \frac{D_{\rm low}-d}{\sigma_{\rm low}}\right)^2\right],
\end{equation}
where $d$ is the actual distance of the gas and hence the actual lower
limit on the pulsar distance. This implies that for any
pulsar distance $D$ we must have $D \geq d$. Hence
we derive the probability distribution for the pulsar's distance as
\begin{equation}
p\left( D \vert D_{\rm low}\right) = \int_0^{\infty} p\left( D \vert
d\right) p\left( d \vert D_{\rm low}\right) {\rm d}d,
\end{equation}
in which
\begin{equation}
p\left( D \vert d \right) \propto H\left( D - d \right)
\end{equation}
with $H(x)$ the heaviside function, as defined above. 
We therefore have
\begin{equation}
p\left( D \vert D_{\rm low}\right) \propto
\int_0^{\infty}H\left( D - d \right) p\left(d \vert D_{\rm low}\right)
    {\rm d}d
= \int_0^D p\left( d \vert D_{\rm  low}\right) {\rm d}d,
\end{equation}
which results in
\begin{equation}\label{eq:HIlowTerm}
p\left( D \vert D_{\rm low} \right) \propto \frac{1}{2} \left[ {\rm erf} \left(
  \frac{ D_{\rm low}}{\sqrt{2} \sigma_{\rm low}} \right) - {\rm erf}
  \left( \frac{ D_{\rm low} - D}{\sqrt{2} \sigma_{\rm low}} \right)
  \right], 
\end{equation}
with ${\rm erf}(x) = \frac{2}{\sqrt{\pi}} \int_0^x
{\rm e}^{-t^2} {\rm d}t$ the error function. 

Analogous to the above derivation, we have the probability
distribution for the distance of the nearest gas not seen in
absorption
\begin{equation}
p\left( d \vert D_{\rm up}\right) \propto
\frac{1}{\sqrt{2\pi}
    \sigma_{\rm up}} \exp\left[ -\frac{1}{2} \left( \frac{D_{\rm up} -
    d}{\sigma_{\rm up}}\right)^2\right];
\end{equation}
which is used in the probability distribution for the pulsar's
distance as:
\begin{equation}
p\left( D \vert D_{\rm up}\right) = \int_0^{\infty} p\left( D \vert d
\right) p\left( d \vert D_{\rm up} \right) {\rm d}d
\end{equation}
with
\begin{equation}
p\left(D\vert d\right) \propto H\left( d - D\right),
\end{equation}
hence:
\begin{equation}
p\left( D \vert D_{\rm up}\right) \propto \int_0^{\infty} H\left( d -
D\right) p\left( d \vert D_{\rm up} \right) {\rm d}d =
\int_D^{\infty}p\left(d\vert D_{\rm up}\right) {\rm d}d
\end{equation}
which results in
\begin{equation}\label{eq:HIupTerm}
p\left( D \vert D_{\rm up} \right) \propto \frac{1}{2} \left[ {\rm erf}
  \left( \frac{D_{\rm up}-D}{\sqrt{2}\sigma_{\rm up}} \right) +
  1\right]. 
\end{equation}

\subsection{The Galactic (``Volumetric'') Term, $p\left( D \vert
  G_{\rm l}, G_{\rm b}\right)$}\label{sss:Vol} 
As derived by \citet{lfl+06}, the distribution of pulsars in the
Galaxy is not homogeneous, but rather follows a distribution of the
form
\begin{equation}
\rho\left( R, \psi, z \right) = \frac{N}{V} \propto
R^B \exp\left[ -\frac{\vert
    z\vert}{E}-C\frac{R-{\rm R_0}}{\rm R_0} \right] {\rm kpc}^{-3},
\end{equation}
with $N$ the number of pulsars per volume $V$ and constants R$_0 =
8.5\,$kpc, $B = 1.9$, $C = 5$ and $E = 330\,$pc for common pulsars and
$E = 500\,$pc for millisecond pulsars \citep[constants from model fit
  C and equations 10 and 11 from][]{lfl+06}.

Since the volume density is invariant with the coordinate system used,
we can use an Earth-based coordinate system based on the Galactic
coordinates of the pulsar and its distance to the Earth, $\left( D,
G_{\rm b}, G_{\rm l}\right)$, for which $\rho\left(D, G_{\rm b},
G_{\rm l}\right) = \rho\left(R, \psi, z\right)$. For the Earth-based
observer the infinitesimal sample volume now becomes
\begin{equation}
\delta V = D^2 \delta D \delta \Omega
\end{equation}
for a pulsar at given distance $D$ and an infinitesimal solid angle
$\delta \Omega$. The number of pulsars in this volume is, hence,
\begin{equation}
\delta N = \rho\left( D, G_{\rm b}, G_{\rm l}\right) D^2 \delta D
\delta \Omega.
\end{equation}
Since the infinitessimal probability $\delta P$ scales with $\delta
N$, we get
\begin{equation}
p\left( D \vert G_{\rm b}, G_{\rm l}\right) \propto 
\rho\left( D, G_{\rm b},G_{\rm l}\right) D^2.
\end{equation}
Consequently, we derive
\begin{equation}\label{eq:VolTerm}
  p\left( D \vert G_{\rm b}, G_{\rm l}\right) \propto R^{1.9} \exp
  \left[ - \frac{\vert z \vert}{E} - 5 \frac{R-{\rm R_0}}{\rm
      R_0}\right]D^2
\end{equation}
with
\begin{equation}\label{eq:zz}
  z\left( D, G_{\rm b} \right) = D \sin{G_{\rm b}}
\end{equation}
and
\begin{equation}\label{eq:RR}
  R\left( D, G_{\rm b}, G_{\rm l} \right) = \sqrt{ {\rm R}_0^2 +
    \left( D \cos{G_{\rm b}}\right)^2 - 2 {\rm R}_0 D \cos{G_{\rm b}}
    \cos{G_{\rm l}}}.
\end{equation}

\subsection{The Pulsar Luminosity Term, $p\left(D \vert S\right)$}
\label{sss:Lum}
Finally, since the radio flux, $S$, of pulsars is related to the
luminosity\footnote{Note that \citet{fk06} define a
  ``pseudo-luminosity'' $L = S D^2$ that avoids the complexities of
  emission beam and viewing geometries. This approach is practical for
  our purposes and hence we copy their usage of $L$ as an effective
  'pseudo-luminosity'.}, $L$, of the pulsar through $SD^2 = L$, this
measure can be used to constrain the pulsar distance, through the
luminosity distribution of radio pulsars derived by
\citet{fk06}. Considering pulsar luminosities at 1.4\,GHz observing
frequency with luminosity expressed in units of mJy\,kpc$^2$, they
proposed a log-normal function with mean $\langle\lambda\rangle =
\langle \log( L ) \rangle = -1.1$ and standard deviation
$\sigma_{\lambda} = 0.9$:
\begin{equation}
p\left( \lambda \right) \propto \exp\left[ -\frac{1}{2} \left(
  \frac{\lambda + 1.1}{0.9}\right)^2\right]. 
\end{equation}
With $\lambda = \log L  = \log S + 2 \log D$, we get
\begin{equation}
p\left( D \right) \propto \left| \frac{\partial \lambda}{\partial D}
\right| p\left( \lambda \right) \propto \frac{1}{D} \exp \left[
  -\frac{1}{2}\left( \frac{\lambda + 1.1}{0.9}\right)^2\right],
\end{equation}
or, given $S$,
\begin{equation}\label{eq:LumTerm}
  p\left( D \vert S \right) \propto \frac{1}{D} \exp \left[
    -\frac{1}{2} \left( \frac{\log{S} + 2 \log{D} + 1.1}{0.9}\right)^2
    \right]. 
\end{equation}
Note that this probability is based on the measured radio flux $S$ of
the \emph{pulsar}, not on the \hi\ flux or the luminosity of an
associated supernova remnant or the like. Also, given the analysis of
\citet{fk06} who derived the luminosity distribution that we use, the
above analysis does not hold for non-radio or bursting pulsars.

\subsection{Combined Distance Probability}
Combining equations \ref{eq:pxTerm}, \ref{eq:HIlowTerm},
\ref{eq:HIupTerm}, \ref{eq:VolTerm} and \ref{eq:LumTerm} into
Equation~\ref{eq:Basic}, we obtain the complete formula for the
pulsar distance given the five measurements listed at the start of
this section:
\begin{align}\label{eq:Final}
  p\left( D \vert \varpi_{\rm meas}, D_{\rm low}, D_{\rm up}, S,
  G_{\rm l}, G_{\rm b}\right)  \propto & \nonumber\\
  & \frac{1}{D^2} \exp\left[ - \frac{1}{2} \left( \frac{ \varpi_{\rm
        meas} - 1/D} {\sigma_{\varpi}}\right)^2\right] \nonumber\\
  & \times \frac{1}{2} \left[ {\rm erf}\left( \frac{ D_{\rm low}} {
      \sqrt{2} \sigma_{\rm low}} \right) - {\rm erf} \left(
    \frac{D_{\rm low} - D}{\sqrt{2} \sigma_{\rm low}}\right)\right]
  \nonumber \\
  & \times \frac{1}{2} \left[ 1 + {\rm erf} \left( \frac{ D_{\rm up} -
      D} { \sqrt{2} \sigma_{\rm up}}\right) \right] \nonumber \\
  & \times R^{1.9} D^2 \exp\left[ - \frac{\left| D \sin{G_{\rm b}}
      \right|}{E} - 5 \frac{R-{\rm R}_0}{\rm R_0}\right] \nonumber \\
  & \times \frac{1}{D} \exp\left[-\frac{1}{2} \left( \frac{ \log{S} +
      2 \log{D} + 1.1}{0.9}\right)^2\right],
\end{align}
with $R$ as in Equation~\ref{eq:RR}. Note that the parallax term
should be replaced by Equation~\ref{eq:pxAsymTerm} in case of
asymmetric uncertainties. (Technically Equation~\ref{eq:pxAsymTerm}
can be applied generally to both asymmetric and symmetric cases, but
for reasons of clarity, we present the more common, simplified formula
here.) Note also that in case measurements are not available, the
relevant terms should be omitted, as $p\left( D\vert \varpi_{\rm
  meas}\right)$ (for example) is nonsensical in the absence of a
$\varpi_{\rm meas}$ measurement.

Equivalently, we find for the pulsar's parallax
\begin{align}\label{eq:PxFinal}
  p\left( \varpi \vert \varpi_{\rm meas}, D_{\rm low}, D_{\rm up}, S, G_{\rm
    l}, G_{\rm b}\right) & \propto \left|\frac{\partial
    D}{\partial\varpi}\right|^5 p\left( D \vert \varpi_{\rm meas}, D_{\rm
    low}, D_{\rm up}, S, G_{\rm l}, G_{\rm b}\right) \nonumber\\
  & \propto \exp\left[-\frac{1}{2} \left(
    \frac{ \varpi_{\rm meas} - \varpi}{\sigma_{\varpi}}\right)^2
    \right]\nonumber\\
  & \phantom{\propto} \times \frac{1}{2\varpi^2} \left[ {\rm erf}\left( \frac{
      D_{\rm low}} { \sqrt{2} \sigma_{\rm low}} \right) - {\rm erf} \left(
    \frac{ D_{\rm low} - 1/\varpi}{\sqrt{2} \sigma_{\rm low}}\right)\right]
  \nonumber \\
  & \phantom{\propto}\times \frac{1}{2\varpi^2} \left[ 1 + {\rm erf} \left(
    \frac{ D_{\rm up} - 1/\varpi} { \sqrt{2} \sigma_{\rm up}}\right)
    \right] \nonumber \\ 
  & \phantom{\propto}\times \frac{R^{1.9}}{\varpi^4} \exp\left[ -
    \frac{\left| \sin{G_{\rm b}}\right|}
         {\varpi E} - 5 \frac{R-{\rm R}_0}{\rm R_0}\right] \nonumber \\
  & \phantom{\propto}\times \frac{1}{\varpi} \exp\left[-\frac{1}{2} \left(
    \frac{ \log{S} - 2 \log{\varpi} + 1.1}{0.9}\right)^2\right],
\end{align}
where each term contributes a factor $\varpi^{-2}$, since
$p(\varpi\vert \varpi_{\rm meas})\delta\varpi = p(D\vert\varpi_{\rm
  meas})\delta D$, $p(\varpi\vert D_{\rm low})\delta\varpi = p(D\vert
D_{\rm low}) \delta D$ etc., implying that each of the five terms
contributes a $\delta D/\delta\varpi$ term.

\begin{figure}
\plotone{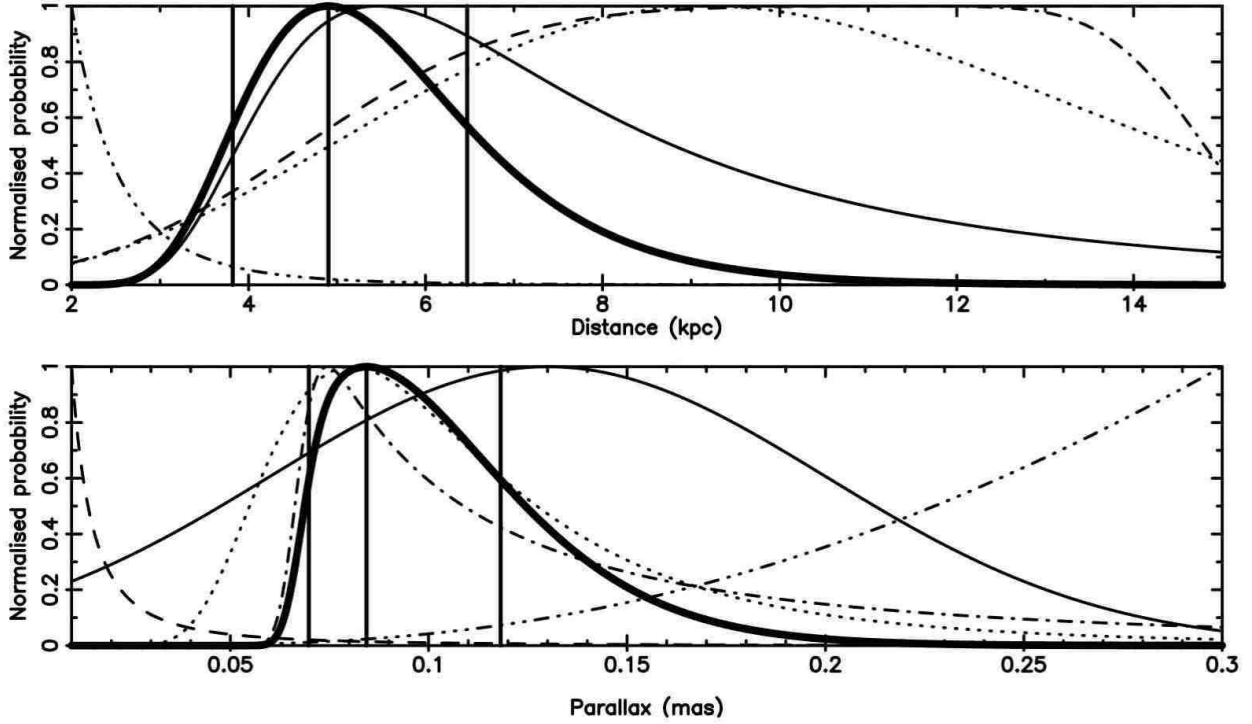}
 \caption{Example output from our likelihood analysis. For
   PSR~J1939+2134, we show the peak-normalised probability
   distributions of the volumetric (dotted) and luminosity
   (triple-dot-dashed) terms, as well as the distance limits from
   \hi\ estimates (dashed and dot-dashed), the parallax measurement
   published by \citet[thin full line]{vbc+09} and the final
   probability distribution for the pulsar distance (thick full line),
   with peak and $1\,\sigma$ uncertainty interval indicated by the
   vertical lines. The top figure shows these distributions as a
   function of distance, while the bottom figure shows the same
   distributions as a function of parallax. Note that, because of the
   non-linear relationship between parallax and distance, the most
   likely distance is not necessarily equal to the inverse of the most
   likely parallax, although these values do converge for small
   uncertainties.\label{fig:example}}
\end{figure}

Equation~\ref{eq:Final} presents the analytic result to the question
first discussed by \citetalias{vlm10}. However, because in that
previous paper parts of the analysis were performed by Monte-Carlo
simulation, our present results are more accurate; and in contrast to
the analysis by \citetalias{vlm10}, which only considered parallax, we
now derive the full formulae for both distance and parallax. We
therefore present in Table~\ref{tbl:vlm10} the bias-corrected parallax
and distance values for the pulsars with parallax measurements first
collated by \citetalias{vlm10}. Results for pulsars with \hi\ distance
limits \citepalias[which were not included in][]{vlm10}, are presented
in Table~\ref{tbl:dist}. For the data in these tables, we use the
following definitions: the corrected distance ($D_{\rm Corr}$) is the
distance for which Equation~\ref{eq:Final} reaches a maximum; for
corrected parallax ($\varpi_{\rm Corr}$) the same convention is used,
based on Equation~\ref{eq:PxFinal}. The 1 $\sigma$ uncertainty
intervals are defined \citep[consistent with][]{cbv+09} as the
narrowest interval that contains 68\% of the integrated probability
density. In practice this means that a level $P^{\ast}$ is found so
that the integral of $p(D)$ for those values of $D$ where $p(D) >
P^{\ast}$, contains 68\% of the total probability. For bimodal
distributions (which only occur towards the Galactic centre, and
particularly for the measurements for PSR~J1752$-$2806), this may
result in two separate regions (a global optimum and a secondary
optimum) which \emph{in combination} contain 68\%
probability. Estimation of these quantities is analytically unfeasible
and is therefore performed numerically. The code used to calculate the
bias-corrected parallax and distance values and uncertainties listed
in the tables, is available as a supplement to this paper and through
an on-line interface on http://psrpop.phys.wvu.edu/LKbias. An example
of the graphical output, showing all five probability terms for
PSR~J1939+2134 (B1937+21), is shown in Figure~\ref{fig:example}.

\section{Discussion and Conclusions}\label{sec:results}
Of the 80 pulsars with \hi\ distance limits, all but one have
post-correction distances consistent (at the 1$\,\sigma$ level
assuming the uncertainties derived from our analysis) with the
\hi\ limits published. The exception is PSR~J2018+2839 (PSR~B2016+28),
which has a lower distance limit of $3.2\pm 2.1$\,kpc, but a parallax
measurement of $1.03\pm 0.10$\,mas \citep{bbgt02}, which dominates the
result and therefore makes the \hi\ distance limit
irrelevant. Furthermore, there is a single source that is beyond the
upper \hi\ distance limit (though within 1$\,\sigma$): this is XTE
J1810-197, for which we determine a bias-corrected distance of $3.7\pm
0.5$\,kpc, which is just beyond the upper distance limit of $3.4\pm
0.6$\,kpc derived from \hi\ observations. Since for this neutron star
both the lower and the upper limit are equal; and because no radio
luminosity is available, the volumetric term determines the slightly
higher distance. For 20 sources, the bias-corrected distance is closer
than the lower \hi\ distance limit (though within 1$\,\sigma$) and 59
(or three out of four) sources are completely within the distance
limits, with typically bias-corrected distances close to the lower
\hi\ distance limit. The fact that our analysis finds sources are more
likely to be closer to the lower rather than upper \hi\ distance limit
(or, indeed, closer even than the lower limit), is unexpected when
seen from the perspective presented by \citet{lk73}. There are two
reasons for this.

First, the upper \hi\ distance limits are mostly past the tangent
point. This means that the volumetric term peaks within -- or close to
-- the range allowed by the \hi\ limits, which causes the volumetric
bias to be either very weak or non-existent. Second, the pulsars to
which \hi\ distance limits have been measured, are mostly bright
sources, with the exception of the flaring neutron stars and those
neutron stars that have \hi\ limits derived from associations with
supernova remnants. The brightness of these pulsars implies a
luminosity term that peaks at very small distances.

Comparing the results in Table~\ref{tbl:dist} and the discussion above
with the results in Table~\ref{tbl:vlm10}, it is clear that the types
of neutron star distance estimates (parallax and \hi\ measurements)
suffer from different statistical biases, although the magnitude of
the biases is limited in both cases. While parallax measurements are
typically biased towards smaller distances (i.e. the sources are
actually \emph{further} away than suggested by the measurement)
because of the relatively limited distance to which this technique
works (and the consequential strong effect of the volumetric term),
the \hi\ measurements are typically biased towards larger distances
(i.e. the sources are often \emph{closer} than suggested by the
measurement) because the volumetric term has little impact and the
luminosity term dominates the analysis.

Finally, of the eight pulsars with both \hi\ distance limits and
parallax distances, only PSR J1857+0943 (B1855+09) has a
bias-corrected parallax that is \emph{in}consistent with the parallax
measurement. The published value of $1.1\pm 0.2$\,mas \citep{vbc+09}
is found to be considerably larger than the most likely value of
$0.6^{+0.2}_{-0.1}$\,mas, which is partly because of the volumetric
information \citepalias[as already found by][who derived a value of $0.9\pm
  0.2$\,mas]{vlm10}, but also because of the \hi\ limits, which place
the pulsar well beyond 1\,kpc. 

\acknowledgements{JPWV is supported by the European Union under Marie
  Curie Intra-European Fellowship 236394. JMW and AAC are supported by
  NSF Grant 0807556. KL is supported by ERC Advanced Grant ``LEAP''
  (Grant Agreement Number 227947, PI Kramer). DRL is supported by the
  West Virginia EPSCoR program and the Research Corporation for
  Scientific Advancement. We gratefully acknowledge use of the ATNF
  Pulsar Catalogue\footnote{\citet{mhth05}; current online version at
    \url{http://www.atnf.csiro.au/research/pulsar/psrcat/}} for the
  determination of basic pulsar parameters. The authors thank Lucas
  Guillemot for useful comments on the draft and Peter den Hartog for
  interesting discussions.}

\bibliographystyle{apj}


\newpage
\begin{deluxetable}{lllccccccc}
\rotate
\tabletypesize{\footnotesize}
\setlength{\tabcolsep}{0.04in}
\tablecolumns{10}
\tablecaption{Pulsar \hi\ distance limits and Lutz-Kelker-bias corrected distances and parallaxes.}
\tablehead{
\multicolumn{2}{c}{Pulsar name} & \colhead{Association} &
\colhead{$\varpi_{\rm meas}$} & \colhead{$D_{\rm low}$} &
\colhead{$D_{\rm up}$} & \colhead{$S_{\rm 1400}$} & \colhead{$\varpi_{\rm Corr}$} & \colhead{$D_{\rm Corr}$} & \colhead{Ref\tablenotemark{e}}\\
\colhead{J2000} & \colhead{B1950} &                   & \colhead{(mas)}    & \colhead{(kpc)}     & \colhead{(kpc)}    & \colhead{(mJy)}          & \colhead{(mas)}              & \colhead{(kpc)}             & 
}
\startdata
J0141+6009       & B0138+59 & \dotfill         &  -      & $2.6\pm 0.7$ & $2.9\pm 0.7$  &   4.5   & $0.30^{+0.07}_{-0.05}$ & $2.3\pm 0.7$        & (1) \\
J0332+5434       & B0329+54 & \dotfill   & $0.94\pm 0.11$& $1.7\pm 0.7$ & $2.0\pm 0.8$  & 203  & $0.8\pm 0.1$           & $1.0\pm 0.1$           & (1, 21) \\
J0358+5413       & B0355+54 & \dotfill   & $0.91\pm 0.16$& $1.4\pm 0.7$ & $2.2\pm 0.9$  &  23  & $0.7\pm 0.2$           & $1.0^{+0.2}_{-0.1}$    & (1, 22) \\
J0738$-$4042     & B0736$-$40 & \dotfill       &  -      & $2.1\pm 0.6$ &      -        &  80     & $0.3\pm 0.1$           & $1.6\pm 0.8$        & (2) \\
J0742$-$2822     & B0740$-$28 & \dotfill       &  -      & $2.0\pm 0.6$ & $6.9\pm 0.8$  &  15     & $0.16^{+0.07}_{-0.03}$ & $2.0^{+1.0}_{-0.8}$ & (3) \\

J0837$-$4135     & B0835$-$41 & \dotfill       &  -      & $1.8\pm 0.8$ & $6.0\pm 0.7$  &  16     & $0.18^{+0.06}_{-0.03}$ & $1.5^{+1.2}_{-0.9}$ & (2) \\
J0908$-$4913     & B0906$-$49 & \dotfill       &  -      & $2.4\pm 1.6$ & $6.7\pm 0.7$  &  10.0   & $0.16^{+0.05}_{-0.02}$ & $1.0^{+1.7}_{-0.7}$ & (3) \\
J0942$-$5552     & B0940$-$55 & \dotfill       &  -      &   -          & $7.5\pm 0.7$  &  10.0   & $0.16^{+0.11}_{-0.03}$ & $0.3^{+0.8}_{-0.2}$ & (2) \\
J1001$-$5507     & B0959$-$54 & \dotfill       &  -      &   -          & $6.9\pm 0.7$  &   6.3   & $0.16^{+0.08}_{-0.03}$ & $0.3^{+1.1}_{-0.3}$ & (3) \\
J1048$-$5832     & B1046$-$58 & \dotfill       &  -      & $2.5\pm 0.5$ & $5.6\pm 0.8$  &   6.5   & $0.18^{+0.05}_{-0.03}$ & $2.9^{+1.2}_{-0.7}$ & (2) \\

J1056$-$6258     & B1054$-$62 & \dotfill       &  -      & $2.5\pm 0.5$ & $2.9\pm 0.5$  &  21     & $0.33^{+0.06}_{-0.05}$ & $2.4\pm 0.5$        & (3) \\
J1124$-$5916     & \dotfill & SNR G292.0+1.8 core &  -   & $3.2\pm2.0$  &      -        &   0.08  & $0.08^{+0.04}_{-0.02}$ & $5^{+3}_{-2}$       & (4)\tablenotemark{a} \\
J1141$-$6545     & \dotfill & \dotfill         &  -      & $3.7\pm 1.7$ &      -        &   3.3   & $0.12^{+0.06}_{-0.04}$ & $3\pm 2$            & (5)\tablenotemark{a} \\
J1157$-$6224     & B1154$-$62 & \dotfill       &  -      & $3.8\pm 1.4$ & $9.0\pm 0.6$  &   5.9   & $0.12^{+0.03}_{-0.01}$ & $4\pm 2$            & (2) \\
J1224$-$6407     & B1221$-$63 & \dotfill       &  -      & $4.3\pm 1.4$ & $11.4\pm 0.7$ &   3.9   & $0.10^{+0.03}_{-0.01}$ & $4\pm 2$            & (2) \\

J1243$-$6423     & B1240$-$64 & \dotfill       &  -      & $4.5\pm 1.9$ & $11.5\pm 0.7$ &  13     & $0.10^{+0.04}_{-0.01}$ & $2\pm 2$            & (1) \\
J1326$-$5859     & B1323$-$58 & \dotfill       &  -      & $3.0\pm 1.0$ &      -        &   9.9   & $0.12^{+0.07}_{-0.04}$ & $3^{+2}_{-1}$       & (6) \\
J1327$-$6222     & B1323$-$62 & \dotfill       &  -      & $5.1\pm 1.7$ & $11.8\pm 0.6$ &  16.0   & $0.093^{+0.031}_{-0.009}$ & $4\pm 2$         & (1) \\
J1359$-$6038     & B1356$-$60 & \dotfill       &  -      & $5.6\pm 1.7$ &      -        &   7.6   & $0.09^{+0.04}_{-0.03}$ & $5\pm 2$            & (1) \\
J1401$-$6357     & B1358$-$63 & \dotfill       &  -      & $1.6\pm 0.5$ & $2.7\pm 0.7$  &   6.2   & $0.31^{+0.09}_{-0.06}$ & $1.8^{+0.7}_{-0.6}$ & (2) \\

J1453$-$6413     & B1449$-$64 & \dotfill       &  -      & $2.5\pm 0.5$ &      -        &  14.0   & $0.13^{+0.09}_{-0.04}$ & $2.8^{+1.3}_{-0.8}$ & (3) \\
J1513$-$5908     & B1509$-$58 & SNR G320.4-01.2 &  -     & $3.8\pm 0.5$ & $6.6\pm 1.4$  &   0.94  & $0.14^{+0.04}_{-0.02}$ & $4.4^{+1.3}_{-0.8}$ & (7) \\
J1559$-$4438     & B1556$-$44 & \dotfill &$0.384\pm 0.081$& $2.0\pm 0.5$&      -        &  40     & $0.32^{+0.07}_{-0.08}$ & $2.3^{+0.5}_{-0.3}$ & (3, 23) \\
J1600$-$5044     & B1557$-$50 & \dotfill       &  -      & $6.4\pm 0.5$ & $18.2\pm 1.2$ &  17.0   & $0.08^{+0.03}_{-0.02}$ & $6.9^{+1.9}_{-0.9}$ & (8) \\
J1602$-$5100     & B1558$-$50 & \dotfill       &  -      & $7.4\pm 0.5$ & $9.4\pm 0.4$  &   5.7   & $0.113^{+0.013}_{-0.007}$ & $8.0^{+0.9}_{-0.7}$ & (8) \\

J1644$-$4559     & B1641$-$45 & \dotfill       &  -      & $4.2\pm 0.3$ & $5.0\pm 0.3$  & 310     & $0.21\pm 0.02$         & $4.5\pm 0.4$        & (1) \\
J1651$-$4246     & B1648$-$42 & \dotfill       &  -      & $4.8\pm 0.3$ &      -        &  16.0   & $0.08^{+0.04}_{-0.02}$ & $5.2^{+2.1}_{-0.6}$ & (9) \\
J1707$-$4053     & B1703$-$40 & \dotfill       &  -      & $3.8\pm 0.5$ &      -        &   7.2   & $0.08^{+0.04}_{-0.02}$ & $4^{+2}_{-1}$       & (9) \\
J1709$-$4429     & B1706$-$44 & \dotfill       &  -      & $2.4\pm 0.6$ & $3.2\pm 0.4$  &   7.3   & $0.31^{+0.05}_{-0.04}$ & $2.6^{+0.5}_{-0.6}$ & (3) \\
J1721$-$3532     & B1718$-$35 & \dotfill       &  -      & $4.4\pm 0.5$ & $5.2\pm 0.6$  &  11.0   & $0.19\pm 0.02$         & $4.6\pm 0.6$        & (9) \\

J1740$-$3015     & B1737$-$30 & \dotfill       &  -      &      -       & $5.5\pm 0.6$  &   6.4   & $0.20^{+0.07}_{-0.03}$ & $0.4^{+1.7}_{-0.3}$ & (8) \\
J1745$-$3040     & B1742$-$30 & \dotfill       &  -      &      -       & $5.5\pm 0.6$  &  13.0   & $0.20^{+0.08}_{-0.03}$ & $0.2^{+1.1}_{-0.2}$ & (8) \\
J1752$-$2806     & B1749$-$28 & \dotfill       &  -      & $0.125\pm 0.025$ &  -        &  18.0   & $0.08\pm 0.03$\tablenotemark{d}             & $0.2^{+1.1}_{-0.1}$ & (1, 10) \\
J1801$-$2304     & B1758$-$23 & \dotfill       &  -      & $3.5\pm 0.9$ & $6.9\pm 0.1$  &   2.2   & $0.149^{+0.033}_{-0.005}$ & $4\pm 1$         & (11) \\
J1803$-$2137     & B1800$-$21 & SNR G8.7-0.1   &  -      & $4.0\pm 0.6$ & $4.9\pm 0.3$  &   7.6   & $0.21\pm 0.02$         & $4.4^{+0.5}_{-0.6}$ & (1) \\
                                                           
J1807$-$0847     & B1804$-$08 & \dotfill       &  -      & $1.5\pm 0.7$ &      -        &  15.0   & $0.11^{+0.10}_{-0.03}$ & $1.5^{+1.2}_{-0.9}$ & (1) \\
\dotfill         & SGR 1806$-$20 & radioflare~2005 &  -  & $6.2\pm 0.1$ &      -        &    -    & $0.06^{+0.02}_{-0.01}$ & $13^{+4}_{-3}$      & (12) \\
XTE J1810$-$197  & \dotfill & radioflare~2006  &  -      & $3.4\pm 0.6$ & $3.4\pm 0.6$  &    -    & $0.25^{+0.04}_{-0.03}$ & $3.6\pm 0.5$        & (13)\tablenotemark{f} \\
J1820$-$0427     & B1818$-$04 & \dotfill       &  -      &      -       & $1.6\pm 0.5$  &   6.1   & $0.5^{+0.2}_{-0.1}$    & $0.3^{+0.6}_{-0.2}$ & (1) \\
J1823+0550       & B1821+05 & \dotfill         &  -      & $1.6\pm 0.5$ &      -        &   1.7   & $0.13^{+0.09}_{-0.04}$ & $2.0^{+1.3}_{-0.8}$ & (1) \\
                                                                 
J1824$-$1945     & B1821$-$19 & \dotfill       &  -      & $3.2\pm 0.5$ &      -        &   4.9   & $0.09^{+0.07}_{-0.02}$ & $3.7^{+1.6}_{-0.9}$ & (8) \\
J1825$-$0935     & B1822$-$09 & \dotfill       &  -      &      -       & $1.9\pm 0.4$  &  12.0   & $0.5^{+0.2}_{-0.1}$    & $0.3^{+0.7}_{-0.2}$ & (8) \\
J1832$-$0827     & B1829$-$08 & \dotfill       &  -      & $4.7\pm 0.3$ & $5.8\pm 0.3$  &   2.1   & $0.18^{+0.02}_{-0.01}$ & $5.2^{+0.5}_{-0.4}$ & (1) \\
J1833$-$0827     & B1830$-$08 & \dotfill       &  -      & $4.0\pm 0.4$ & $5.3\pm 0.3$  &   3.6   & $0.20\pm 0.02$         & $4.5\pm 0.5$        & (9) \\
J1833$-$1034     & \dotfill & SNR G21.5-0.9    &  -      & $4.0\pm 0.3$ & $4.1\pm 0.3$  &   0.071 & $0.24\pm 0.02$         & $4.1\pm 0.3$        & (14)\tablenotemark{a,g} \\
                                                           
AXP 1E1841$-$045 & \dotfill & SNR Kes73        &  -      & $7.5\pm 1.0$ & $10.2\pm 0.3$ &   -    & $0.102^{+0.012}_{-0.005}$ & $9.6^{+0.6}_{-1.4}$ & (15)\tablenotemark{b,f} \\
J1846$-$0258     & \dotfill & SNR Kes75        &  -      & $5.5\pm 0.4$ & $5.9\pm 0.5$  &    -    & $0.17\pm 0.01$         & $5.8^{+0.5}_{-0.4}$ & (16)\tablenotemark{a,g} \\
J1848$-$0123     & B1845$-$01 & \dotfill       &  -      & $4.2\pm 0.4$ & $4.8\pm 0.4$  &   8.6   & $0.21\pm 0.02$         & $4.4\pm 0.4$        & (1) \\
J1852+0031       & B1849+00 & \dotfill         &  -      & $7.1\pm 1.2$ & $16.6\pm 0.9$ &   2.2   & $0.070^{+0.025}_{-0.009}$ & $8\pm 2$         & (1) \\
J1857+0212       & B1855+02 & \dotfill         &  -      & $6.9\pm 1.3$ &      -        &   1.6   & $0.08^{+0.03}_{-0.02}$ & $8\pm 2$            & (1) \\
                                                           
J1857+0943       & B1855+09 & \dotfill    & $1.1\pm 0.2$ & $1.6\pm 0.5$ & $2.0\pm 0.4$  &   5     & $0.6^{+0.2}_{-0.1}$    & $0.9\pm 0.2$        & (1, 24) \\
J1901+0331       & B1859+03 & \dotfill         &  -      & $6.8\pm 1.4$ & $15.1\pm 0.7$ &   4.2   & $0.075^{+0.027}_{-0.008}$ & $7\pm 2$         & (1) \\
J1901+0716       & B1859+07 & \dotfill         &  -      & $2.8\pm 0.5$ & $4.7\pm 0.8$  &   0.9   & $0.20^{+0.05}_{-0.03}$ & $3.4^{+0.9}_{-0.7}$ & (1) \\
J1902+0556       & B1900+05 & \dotfill         &  -      & $3.1\pm 0.4$ & $4.3\pm 0.5$  &   1.2   & $0.23^{+0.04}_{-0.03}$ & $3.6^{+0.6}_{-0.5}$ & (1) \\
J1902+0615       & B1900+06 & \dotfill         &  -      & $6.5\pm 1.4$ & $15.8\pm 0.8$ &   1.1   & $0.071^{+0.024}_{-0.007}$ & $7^{+3}_{-2}$    & (1) \\
                                                           
J1903+0135       & B1900+01 & \dotfill         &  -      & $2.8\pm 0.4$ & $4.0\pm 0.4$  &   1.1   & $0.26^{+0.04}_{-0.03}$ & $3.3^{+0.6}_{-0.5}$ & (1) \\
J1906+0641       & B1904+06 & \dotfill         &  -      & $6.5\pm 1.5$ & $14.0\pm 0.5$ &   1.7   & $0.077^{+0.026}_{-0.006}$ & $7\pm 2$         & (1) \\
J1909+0254       & B1907+02 & \dotfill         &  -      & $3.8\pm 0.5$ &      -        &   0.63  & $0.09^{+0.05}_{-0.03}$ & $4.5^{+2.2}_{-0.9}$ & (17) \\
J1909+1102       & B1907+10 & \dotfill         &  -      & $4.3\pm 0.6$ & $6.0\pm 1.6$  &   1.9   & $0.14^{+0.04}_{-0.03}$ & $4.8^{+1.1}_{-0.8}$ & (1) \\
J1915+1009       & B1913+10 & \dotfill         &  -      & $6.0\pm 1.5$ & $14.5\pm 0.8$ &   1.3   & $0.077^{+0.027}_{-0.009}$ & $7\pm 2$         & (1) \\

J1916+1312       & B1914+13 & \dotfill         &  -      & $4.0\pm 0.7$ & $5.7\pm 1.7$  &   1.2   & $0.14^{+0.04}_{-0.03}$ & $4.5^{+1.2}_{-0.9}$ & (1) \\
J1917+1353       & B1915+13 & \dotfill         &  -      & $4.8\pm 1.0$ & $5.7\pm 1.7$  &   1.9   & $0.14\pm 0.03$         & $5\pm 1$             & (1) \\
J1921+2153       & B1919+21 & \dotfill         &  -      &      -       & $2.8\pm 1.2$  &   6     & $0.29^{+0.15}_{-0.08}$ & $0.3^{+0.8}_{-0.2}$ & (1) \\
J1922+2110       & B1920+21 & \dotfill         &  -      & $4.8\pm 1.8$ & $16.2\pm 1.0$ &   1.4   & $0.08^{+0.03}_{-0.02}$ & $4\pm 2$            & (17) \\
J1926+1648       & B1924+16 & \dotfill         &  -      & $5.2\pm 1.8$ & $14.9\pm 0.8$ &   1.3   & $0.075^{+0.028}_{-0.008}$ & $6^{+3}_{-2}$    & (17) \\

J1932+1059       & B1929+10 & \dotfill  & $2.77\pm 0.07$ &      -       & $1.6\pm 0.5$  &  36     & $0.9^{+1.0}_{-0.3}$    & $0.31^{+0.09}_{-0.06}$ & (1, 22) \\
J1932+2020       & B1929+20 & \dotfill         &  -      & $4.8\pm 1.8$ & $14.9\pm 0.9$ &   1.2   & $0.076^{+0.029}_{-0.009}$ & $5^{+3}_{-2}$    & (1) \\
J1930+1852       & \dotfill & PWN G54.1+0.3    &  -      & $5.0\pm 1.8$ & $12.6\pm 0.6$ &   0.06  & $0.085^{+0.020}_{-0.007}$ & $7^{+3}_{-2}$    & (18)\tablenotemark{b,g}\\
J1932+2220       & B1930+22 & \dotfill         &  -      & $10.4\pm 0.6$& $13.7\pm 0.7$ &   1.2   & $0.081^{+0.010}_{-0.007}$ & $10.9^{+1.3}_{-0.8}$ & (1) \\
J1935+1616       & B1933+16 & \dotfill & $0.22^{+0.08}_{-0.12}$&$5.2\pm 1.7$&   -       &  42     & $0.13^{+0.05}_{-0.04}$ & $3.7^{+1.3}_{-0.8}$ & (1, 25) \\

J1939+2134       & B1937+21 & \dotfill  & $0.13\pm 0.07$ & $4.6\pm 1.9$ & $14.8\pm 0.9$ &  10     & $0.08^{+0.03}_{-0.01}$ & $5^{+2}_{-1}$       & (1, 24) \\
J1946+1805       & B1944+17 & \dotfill         &  -      &      -       & $1.9\pm 0.7$  &  10     & $0.4^{+0.2}_{-0.1}$    & $0.3^{+0.6}_{-0.2}$ & (1) \\
J1952+3252       & B1951+32 & SNR CTB80        &  -      & $3.1\pm 2.0$ &      -        &   1.0   & $0.11^{+0.06}_{-0.03}$ & $3\pm 2$       & (19)\tablenotemark{b,h}\\
J2004+3137       & B2002+31 & \dotfill         &  -      & $7.0\pm 0.7$ & $12.0\pm 0.7$ &   1.8   & $0.092^{+0.023}_{-0.009}$ & $8^{+2}_{-1}$    & (1) \\
J2018+2839       & B2016+28 & \dotfill   & $1.03\pm 0.10$ & $3.2\pm 2.1$ &       -       &  30     & $0.9\pm 0.1$           & $0.98^{+0.11}_{-0.09}$ & (1, 21) \\
                                                                  
J2022+2854       & B2020+28 & \dotfill  & $0.37\pm 0.12$ & $3.1\pm 2.1$ &       -       &  38     & $0.22^{+0.10}_{-0.07}$ & $2.1^{+0.6}_{-0.4}$ & (1, 21) \\
J2113+4644       & B2111+46 & \dotfill         &  -      & $4.3\pm 0.8$ & $6.5\pm 0.7$  &  19     & $0.17^{+0.03}_{-0.02}$ & $4\pm 1$            & (1) \\
J2257+5909       & B2255+58 & \dotfill         &  -      & $3.3\pm 0.7$ &      -        &   9.2   & $0.13^{+0.08}_{-0.04}$ & $3\pm 1$            & (1) \\
\dotfill         & AXP 1E2259+586 & SNR CTB109 &  -      & $4.0\pm 0.8$ & $4.0\pm 0.8$\tablenotemark{c} & - & $0.21^{+0.04}_{-0.03}$ & $4.1\pm 0.7$ & (20) \\
J2321+6024       & B2319+60 & \dotfill         &  -      & $2.6\pm 0.6$ &      -        &  12     & $0.13^{+0.10}_{-0.05}$ & $2.7^{+1.2}_{-0.9}$ & (1) \\
\enddata

\tablenotetext{a}{Distance limit {\it{uncertainty}} is derived by the current authors.}
\tablenotetext{b}{Distance limit  {\it{and uncertainty}} are derived by the current authors. See notes in body of paper.}
\tablenotetext{c}{Upper limit based on CO emission from a molecular cloud associated with SNR~CTB109.}
\tablenotetext{d}{PSR~J1752$-$2806 (B1749$-$28) has a secondary optimum in its parallax, at $0.15^{+0.03}_{-0.02}$\,mas.}
\tablenotetext{e}{References: (1) \citet{fw90}; (2) \citet{jkww96}; (3) \citet{kjww95}; (4) \citet{gw03}; (5) \citet{obv02a}; (6) \citet{sdw+96}; (7) \citet{gbm+99}; (8) \citet{jkww01}; (9) \citet{wsfj95}; 
  (10) \citet{jkww01}; (11) \citet{fkv93} and Frail personal communication; (12) \citet{mg05}; (13) \citet{mcr+08}; (14) \citet{tl08b}; (15) \citet{tl08a}; (16) \citet{lt08}; (17) \citet{wsx+08}; (18) \citet{ltw08}; 
  (19) \citet{ss00}; (20) \citet{tll10}; (21) \citet{bbgt02}, (22) \citet{ccv+04}, (23) \citet{dtb09}, (24) \citet{vbc+09}, (25) \citet{cbv+09}.}
\tablenotetext{f}{Original authors cited multiple distances derived from multiple rotation curves; we choose the standard flat model.}
\tablenotetext{g}{Reverted flat rotation model to old Galactic constants $(R_0,\Theta_0)=(8.5\,{\rm kpc},220\,{\rm km}/{\rm s})$}
\tablenotetext{h}{The original authors did not publish the absorption spectrum at velocities enabling a $D_{\rm up}$ measurement.}
\label{tbl:dist}
\end{deluxetable}

\newpage
\begin{deluxetable}{llccccc}
\tabletypesize{\scriptsize}
\setlength{\tabcolsep}{0.04in}
\tablecolumns{7}
\tablecaption{Pulsar parallax measurements and Lutz-Kelker-bias corrected distances and parallaxes for pulsars without \hi\ limits but with parallax measurements only.}
\tablehead{
  \multicolumn{2}{c}{Pulsar name} & \colhead{$\varpi_{\rm meas}$} &
  \colhead{$S_{\rm 1400}$} & \colhead{$\varpi_{\rm Corr}$} &
  \colhead{$D_{\rm Corr}$} & \colhead{Ref.}\\
  \colhead{J2000} & \colhead{B1950} & \colhead{(mas)} &
  \colhead{(mJy)} & \colhead{(mas)} & \colhead{(kpc)} & }
\startdata
J0030+0451   & \dotfill & $3.3\pm 0.9$   &   0.6  & $1.6^{+1.0}_{-0.8}$ & $0.28^{+0.10}_{-0.06}$ & \citet{lkn+06,lzb+00} \\ 
J0034$-$0721 & B0031$-$07 & $0.93^{_+0.08}_{-0.07}$&11 & $0.93^{+0.08}_{-0.07}$ & $1.03\pm 0.08$   & \citet{cbv+09} \\
J0108$-$1431 & \dotfill & $4.2\pm 1.4$   &   1.0  & $1.4^{+1.4}_{-0.7}$ & $0.21^{+0.09}_{-0.05}$ & \citet{dtbr09} \\
J0139+5814 & B0136+57 & $0.37\pm 0.04$ &   4.6  & $0.37\pm 0.04$      & $2.6^{+0.3}_{-0.2}$    & \citet{cbv+09} \\
J0437$-$4715 & \dotfill & $6.396\pm 0.054$ & 142  & $6.39\pm 0.05$    & $0.156\pm 0.001$         & \citet{dvtb08} \\

J0452$-$1759 & B0450$-$18 & $0.64^{+1.4}_{-0.6}$  & 5.3 & $0.7^{+0.6}_{-0.3}$ & $0.4^{+0.2}_{-0.1}$ & \citet{cbv+09} \\
J0454+5543   & B0450+55 & $0.84^{+0.04}_{-0.05}$&13 & $0.84^{+0.04}_{-0.05}$ & $1.18^{+0.07}_{-0.05}$ & \citet{cbv+09} \\
J0538+2817   & \dotfill & $0.72^{+0.12}_{-0.09}$& 1.9 & $0.69^{+0.11}_{-0.09}$ & $1.3\pm 0.2$    & \citet{cbv+09,lwf+04} \\
J0613$-$0200 & \dotfill & $0.80\pm 0.35$ &   1.4  & $0.4^{+0.3}_{-0.2}$ & $0.9^{+0.4}_{-0.2}$    & \citet{vbc+09} \\
J0630$-$2834 & B0628$-$28 & $3.0\pm 0.4$   &  23    & $2.8\pm 0.4$        & $0.32^{+0.05}_{-0.04}$ & \citet{dtbr09} \\

J0633+1746 & \dotfill & $4.0\pm 1.3$   &    -   & $0.2^{+0.5}_{-0.1}$ & $0.25^{+0.23}_{-0.08}$ & \citet{fwa07} \\
J0659+1414 & B0656+14 & $3.47\pm 0.36$ &   3.7  & $3.3\pm 0.4$        & $0.28\pm 0.03$         & \citet{btgg03} \\
J0720$-$3125 & \dotfill & $2.77\pm 0.89$ &    -   & $0.2^{+0.8}_{-0.1}$ & $0.4^{+0.3}_{-0.1}$    & \citet{kva07} \\
J0737$-$3039A& \dotfill & $0.87\pm 0.14$ &   1.6  & $0.80\pm 0.14$      & $1.1^{+0.2}_{-0.1}$    & \citet{dbt09,bjd+06} \\
J0751+1807 & \dotfill & $1.6\pm 0.8$   &   3.23 & $0.6^{+0.6}_{-0.3}$ & $0.4^{+0.2}_{-0.1}$    & \citet{nss+05} \\

J0814+7429 & B0809+74 & $2.31\pm 0.04$ &  10    & $2.31\pm 0.04$      & $0.432^{+0.008}_{-0.007}$ & \citet{bbgt02} \\
J0820$-$1350 & B0818$-$13 & $0.51^{+0.03}_{-0.04}$&7& $0.51^{+0.03}_{-0.04}$ & $1.9\pm 0.1$        & \citet{cbv+09} \\
J0826+2637 & B0823+26 & $2.8\pm 0.6$   &  10    & $2.4\pm 0.6$        & $0.32^{+0.08}_{-0.05}$ & \citet{gtwr86} \\
J0835$-$4510 & B0833$-$45 & $3.5\pm 0.2$   &1100    & $3.5\pm 0.2$        & $0.28\pm 0.02$         & \citet{dlrm03,bf74} \\
J0922+0638 & B0919+06 & $0.82\pm 0.13$ &   4.2  & $0.82^{+0.13}_{-0.12}$ & $1.1^{+0.2}_{-0.1}$ & \citet{ccl+01} \\

J0953+0755 & B0950+08 & $3.82\pm 0.07$ &  84    & $3.82\pm 0.07$      & $0.261\pm 0.005$       & \citet{bbgt02} \\
J1012+5307 & \dotfill & $1.22\pm 0.26$ &   3    & $1.11\pm 0.25$      & $0.7^{+0.2}_{-0.1}$    & \citet{lwj+09} \\
J1022+1001 & \dotfill & $1.8\pm 0.3$   &   3    & $1.7\pm 0.3$        & $0.52^{+0.09}_{-0.07}$ & \citet{vbc+09} \\
J1024$-$0719 & \dotfill & $1.9\pm 0.4$   &   0.66 & $1.5\pm 0.4$        & $0.49^{+0.12}_{-0.08}$ & \citet{hbo06} \\
J1045$-$4509 & \dotfill & $3.3\pm 1.9$   &   3    & $0.3^{+0.4}_{-0.1}$ & $0.23^{+0.17}_{-0.07}$ & \citet{vbc+09} \\

J1136+1551 & B1133+16 & $2.80\pm 0.16$ &  32    & $2.80\pm 0.16$      & $0.35\pm 0.02$         & \citet{bbgt02} \\
J1239+2453 & B1237+25 & $1.16\pm 0.08$ &  10    & $1.16\pm 0.08$      & $0.84+0.06-0.05$       & \citet{bbgt02} \\
J1300+1240 & B1257+12 & $1.3\pm 0.4$   &   2    & $1.0^{+0.4}_{-0.3}$ & $0.6^{+0.2}_{-0.1}$    & \citet{wdk+00} \\
J1456$-$6843 & B1451$-$68 & $2.2\pm 0.3$   &  80    & $2.1\pm 0.3$        & $0.43^{+0.06}_{-0.05}$ & \citet{bmk+90a,mhm80} \\
J1509+5531 & B1508+55 & $0.47\pm 0.03$ &   8    & $0.47\pm 0.03$      & $2.1\pm 0.1$           & \citet{cbv+09} \\

J1537+1155 & B1534+12 & $0.98\pm 0.05$ &   0.6  & $0.97\pm 0.05$      & $1.01\pm 0.05$         & \citet{sttw02} \\
J1543+0929 & B1541+09 & $0.13\pm 0.02$ &   5.9  & $0.16\pm 0.02$      & $5.9^{+0.6}_{-0.5}$    & \citet{cbv+09} \\
J1600$-$3053 & \dotfill & $0.20\pm 0.15$ &   3.2  & $0.21^{+0.10}_{-0.07}$& $2.4^{+0.9}_{-0.6}$  & \citet{vbc+09,jbo+07} \\
J1643$-$1224 & \dotfill & $2.2\pm 0.4$   &   4.8  & $1.9\pm 0.4$        & $0.42^{+0.09}_{-0.06}$ & \citet{vbc+09} \\
J1713+0747 & \dotfill & $0.94\pm 0.05$ &   8    & $0.93\pm 0.05$      & $1.05^{+0.06}_{-0.05}$ & \citet{vbc+09} \\

J1744$-$1134 & \dotfill & $2.4\pm 0.1$   &   3    & $2.4\pm 0.1$        & $0.42\pm 0.02$         & \citet{vbc+09} \\
J1856$-$3754 & \dotfill & $6.2\pm 0.6$   &    -   & $6.0\pm 0.6$        & $0.16^{+0.02}_{-0.01}$ & \citet{vk07} \\
J1900$-$2600 & B1857$-$26 & $0.5\pm 0.6$   &  13    & $0.3^{+0.3}_{-0.1}$ & $0.7^{+0.4}_{-0.2}$    & \citet{fgbc99} \\
J1909$-$3744 & \dotfill & $0.79\pm 0.02$ &   3    & $0.79\pm 0.02$      & $1.26\pm 0.03$         & \citet{vbc+09,jbv+03} \\
J2022+5154   & B2021+51 & $0.50\pm 0.07$ &  27    & $0.49\pm 0.07$      & $1.8^{+0.3}_{-0.2}$    & \citet{bbgt02} \\

J2048$-$1616 & B2045$-$16 &$1.05^{+0.03}_{-0.02}$&13&$1.05^{+0.03}_{-0.02}$&$0.95^{+0.02}_{-0.03}$ & \citet{cbv+09} \\
J2055+3630   & B2053+36 & $0.17\pm 0.03$ &   2.6  & $0.17\pm 0.03$      & $5.0^{+0.8}_{-0.6}$    & \citet{cbv+09} \\
J2124$-$3358 & \dotfill & $3.1\pm 0.6$   &   1.6  & $2.5^{+0.6}_{-0.7}$ & $0.30^{+0.07}_{-0.05}$ & \citet{vbc+09} \\
J2129$-$5721 & \dotfill & $1.9\pm 0.9$   &   1.4  & $0.5^{+0.6}_{-0.3}$ & $0.4^{+0.2}_{-0.1}$    & \citet{vbc+09} \\
J2144$-$3933 & \dotfill & $6.05\pm 0.56$ &   0.8  & $5.77\pm 0.57$      & $0.16^{+0.02}_{-0.01}$ & \citet{dtbr09} \\

J2145$-$0750 & \dotfill & $1.6\pm 0.3$   &   8    & $1.5\pm 0.3$        & $0.57^{+0.11}_{-0.08}$ & \citet{vbc+09} \\
J2157+4017 & B2154+40 & $0.28\pm 0.06$ &  17    & $0.29\pm 0.05$      & $2.9^{+0.5}_{-0.4}$    & \citet{cbv+09} \\
J2313+4253 & B2310+42 & $0.93^{+0.06}_{-0.07}$&15&$0.92^{+0.06}_{-0.07}$& $1.06^{+0.08}_{-0.06}$& \citet{cbv+09}
\enddata
\label{tbl:vlm10}
\end{deluxetable}

\end{document}